# Main-chain Polyimidazolium Polymers by One-pot Synthesis and Application as Nitrogen-doped Carbon Precursors

Konrad Grygiel,[1] Sarah Kirchhecker,[1] Jiang Gong,[1] Markus Antonietti,[1] Davide Esposito,[1] Jiayin Yuan[1,2]*

__________

[1]Dr. K. Grygiel, Dr. Kirchhecker, Dr. J. Gong, Prof. Dr. M. Antonietti, Dr. Esposito, Prof. Dr. J. Yuan

Department of Colloid Chemistry, Max Planck Institute of Colloids and Interfaces, Am Mühlenberg 1, 14476 Potsdam, Germany

E-mail: jiayin.yuan@mpikg.mpg.de

[2]Prof. Dr. J. Yuan

Department of Chemistry and Biomolecular Science, and Center for Advanced Materials Processing, Clarkson University, 8 Clarkson Avenue, 13699 Potsdam, USA

Email jyuan@clarkson.edu

__________

Abstract:

This paper reports on the one-pot synthesis of main-chain imidazolium-containing polymers, some of which show unusually high thermal stability. The imidazolium polymers were obtained by modified Debus-Radziszewski reactions for the chain build-up from simple organic compounds, here pyruvaldehyde, formaldehyde, acetic acid, and a variety of diamines. The reactions were performed in aqueous media at ambient conditions, being synthetically elegant, convenient and highly efficient. Finally, a simple anion-metathesis reaction was conducted to replace the acetate anion with dicyanamide, and the thermal properties of the main-chain polyimidazoliums before and after anion exchange were studied in detail, which demonstrated chain cross-linking by the counterion and a coupled unusually high carbonization yield of up to 66 wt% at 900 °C.

1. **Introduction**

Polymers containing a high density of imidazolium units have recently gained much interest, majorly due to their classic structural ionic unit, well known from ionic liquids [1-4]. A common but non-exclusive method to introduce imidazolium-based ionic monomers into a polymer structure design follows the poly(ionic liquid) route, which synthesizes functional ionic polymers *via* polymerization of ionic liquid monomers. These polymers possess versatile properties that are tunable *via* varying anions without changing the polycation main chain and can serve in a broad range of applications.[5] For example, the solubility, thermal and electrochemical stability, as well as the ionic conductivity of these polymers are all affected by anion exchange.[6-9] In addition, polyimidazolium polymers emerge as efficient stabilizers for omnibus nanomaterials such as carbon nanotubes,[10,11] metal nanoparticles,[12] conductive polymer particles,[13] and cellulose nanofibrils[14] in aqueous and organic media. They have also been used in electrochemical devices, such as dye-sensitized solar cells,[15] field-effect transistors,[16] lithium batteries,[17,18] and light-emitting electrochemical cells.[19]

This traditional method to prepare polyimidazoliums typically requires several synthetic steps. The common total synthesis towards imidazolium-type polymers begins with the Debus-Radziszewski reaction of ammonia, formaldehyde, and a dicarbonyl compound to yield neutral imidazole ring compounds.[20] The as-synthesized imidazoles are subsequently converted into a polymerizable *N*-vinyl imidazole by catalytic addition of acetylene.[21] The synthesis till this step has been commercialized and is well-known in industry. *N*-vinyl imidazole as commercially available source can undergo quaternization reaction with alkyl halides, which is a common method for the synthesis of ionic liquid monomers.[5] Finally, various polymerization techniques, including free radical polymerization, atom transfer radical polymerization,[22,23] cobalt-mediated radical polymerization,[24] or reversible addition-fragmentation transfer polymerization[25,26] can be used to obtain imidazolium-type polymers. On a laboratory scale, imidazolium-containing polymeric networks were also synthesized using a polymer analogous reaction described by Enthaler *et al*. In this synthetic route, tetrakis(4-aminophenyl) methane was reacted with glyoxal in 1,1,1,3,3,3-hexafluoroisopropanol, forming a polyimine network. After supercritical drying, the porous material was subsequently transformed into a polyimidazolium chloride salt in a ring closure reaction with chloromethyl ethyl ether.[27] Analogously, Coskun *et al.* synthesized imidazolium chloride containing networks in two steps, starting from the reaction of aniline derivatives with glyoxal and acetic acid in dioxane, followed by the reaction of its products with formaldehyde in the presence of HCl and in THF as the solvent.[28] Synthetic routes towards ionenes bearing imidazolium cation in the main-chain resemble substantially the polyimidazolium formation in this work [29-31]. Such ionenes can be synthesized in two steps from the imidazole or methylimidazole compounds: the first step is to obtain an imidazole dimer by reacting excessive imidazoles with dibromoalkanes; in the second step the dimer further reacts with dibromoalkane in a polycondensation model to produce ionenes. This two-step process however involves organic solvents, and strict reaction conditions. Moreover, the necessity of using organic solvents and fossil oil-based hazardous chemicals leaves a rather unfavorable eco-footprint on the environment. Driven by the unique properties of the polymers and their high potential value in various practical usages, the low-cost, environment-friendly synthesis of imidazolium polymers is highly demanded.

Recently, we reported an efficient and green method for a series of novel, bifunctional imidazolium compounds obtained by modification of the Debus-Radziszewski reaction.[32] The developed protocol utilizes amino acids and bio-derived carbonyl compounds to produce bi-functional imidazolium zwitterions in high yields.[33] This synthetic method is facile and elegant, since it is not sensitive to oxygen and can be conducted in aqueous media at room temperature. In the present study, this methodology is further extended by substituting diamine compounds for amino acids. It was anticipated that increasing the functionality of amine-containing molecules from 1 to 2 may yield immediately linear imidazolium-type polymers rather than low molecular weight molecules. This modification of the synthetic procedure then opens up a new avenue to synthesize imidazolium-type polymers in one-step from low molecular weight raw resources that can be easily accessed either industrially or from biomass.

2. **Experimental Section**

   **2.1. Materials:**

Acetic acid (purity ≥ 99%, Sigma), ethylenediamine (purity = 99%, Alfa Aesar), 1,4-diaminobutane (purity = 99%, Aldrich), 1,5-diaminopentane (purity = 98%, Acros Organics), 1,6-diaminohexane (purity = 98%, Alfa Aesar), 1,8-diaminooctane (purity = 98%, Aldrich), 1,10-diaminodecane (purity = 97%, Aldrich), 1,12-diaminododecane (purity = 98%, Aldrich), *p*-phenylenediamine (purity ≥ 99%, Aldrich), lithium bis(trifluoromethane sulfonyl)imide (LiTFSI, purity = 99%, Io-li-tec), potassium hexafluorophosphate (purity = 99%, Alfa Aesar), sodium dicyanamide (NaN(CN)$_2$, purity ≥ 97%, Aldrich), formaldehyde (37% aqueous solution, Applichem) and pyruvaldehyde (40% aqueous solution, Sigma) were used without further purification.

### 2.2. General Procedure for Polyimidazolium synthesis

In a typical reaction, MiliQ® water (3.7 mL) and glacial acetic acid (3.3 mL, 57 mmol) were added under vigorous stirring to 1.0 g (9.4 mmol) of cadaverine. This mixture was injected into a mixture of methylglyoxal (1.46 mL, 9.0 mmol) and formaldehyde (0.71 mL, 9.0 mmol), stirred for 24 h, diluted with MiliQ® water and dialyzed against water using dialysis tubing with molecular weight cutoff of 3.5 kD. A similar procedure was applied for other reactions with different diamines (except *p*-phenylenediamine). The reaction with *p*-phenylenediamine was analogous to the above mentioned reactions except that the overall amount of water in the reaction was set to 40 mL and the reaction time was 15 min.

### 2.3. General Procedure for Anion Exchange of Imidazolium Polymers

For the anion exchange from acetate to TFSI, 1.2 molar excess of LiTFSI solution (calculated to the theoretical amount of imidazolium rings at 100% of conversion in the Debus-Radziszewski reaction) in MiliQ® water (20 mL) was added to the dialyzed solution of polymers under stirring. After 30 min the precipitate was centrifuged out and washed 3 times with water. The product was then dried under high vacuum at 80 °C until the weight was constant.

Anion exchange from acetate to $N(CN)_2^-$ anion was performed in an analogous method to the above-mentioned one except that 5.0 molar excess of $NaN(CN)_2$ dissolved in 100 mL of water was used for the reactions.

### 2.4. Characterization

Attenuated Total Reflection Fourier-transform infrared spectroscopy (ATR-FTIR) was performed at room temperature with a BioRad 6000 FT-IR spectrometer equipped with a Single Reflection Diamond ATR.

Differential scanning calorimetry (DSC) measurements were done under nitrogen flow using a Perkin-Elmer DSC-1 instrument.

Gel permeation chromatography (GPC) measurement was performed using a NOVEMA Max linear XL column with a mixture of 80% of acetate buffer and 20% of methanol as the eluent (flow rate = 1.00 mL min$^{-1}$) Pullalan standards, and using an RI-101 Refractometer (Shodex) as the RI detector.

Nuclear Magnetic Resonance (NMR) Spectroscopy: Carbon nuclear magnetic resonance ($^{13}$C-NMR) spectra were recorded at room temperature using a Bruker DPX-400 spectrometer operating at 100.6 MHz. Proton nuclear magnetic resonance ($^{1}$H-NMR) spectra were recorded at room temperature using a Bruker DPX-400 spectrometer operating at 400.1 MHz.

Thermogravimetric analysis (TGA) experiment was performed under nitrogen flow at a heating rate of 10 °C min$^{-1}$ using a Netzsch TG209-F1 apparatus.

Inductively coupled plasma-optical emission spectroscopy (ICP-OES) was carried out using an ICP OES Optima 2100 DV (Perkin Elmer) to measure the metal content of the carbon product.

The surface element composition was characterized by means of X-ray photoelectron spectroscopy (XPS) carried out on a VG ESCALAB MK II spectrometer using Al K$α$ exciting radiation from an X-ray source operated at 10.0 kV and 10 mA.

X-ray diffraction (XRD) pattern was recorded on a Bruker D8 diffractometer using Cu K$α$ radiation ($λ$ = 0.154 nm) and a scintillation counter.

## 3. Results and Discussion

### 3.1. One-pot Synthesis of Main-chain Imidazolium Polymers

In this contribution, we report the one-step synthesis of imidazolium-type polymers from Debus-Radziszewski reactions of pyruvaldehyde, formaldehyde and diamines in aqueous acetic acid solution (Figure 1), which is a part of a PhD thesis conducted in our lab recently.[34] A very recent report by Lindner[35] came to our attention afterwards which also exploits the potential of Debus-Radziszewski reaction in polyimidazolium synthesis. In the present approach, pentamethylene-1,5-diamine (also termed cadaverine), which can be produced from cellobiose, was first applied as a model diamine compound and reacted with acetic acid, pyruvaldehyde and formaldehyde. Polyimidazolium polymers derived from other diamines in similar conditions were also obtained. At the end, the unusually high carbonization yield of polyimidazoliums derived from this synthetic method is discussed.

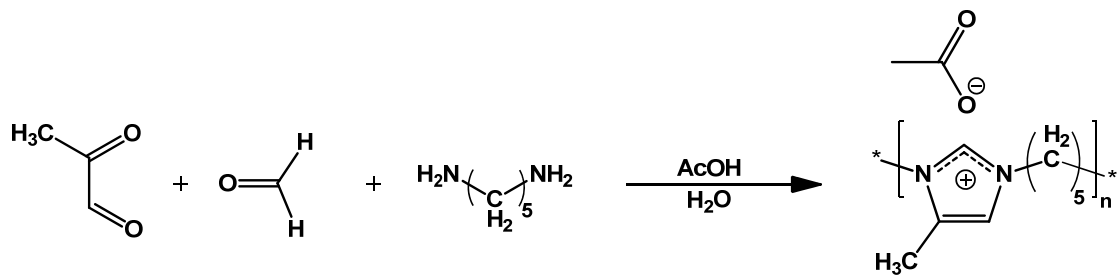

**Figure 1.** Debus-Radziszewski synthesis of main-chain imidazolium polymers.

## 3.2. Synthesis and Characterization of Imidazolium Polymers From Aliphatic Diamines

The developed synthetic method yields in generally light- to dark-brown, water-soluble, polyimidazolium products termed polyCx$^+$Ac$^-$ (*x* denotes the number of carbon atoms in the diamines) in acidic aqueous solution. After dialysis in water, high yields (> 97%) can be determined, which already indicates a sufficiently high molecular weight. The chemical structure of our first polyimidazolium polymer polyC5$^+$Ac$^-$ derived from pentamethylene-1,5-diamine was confirmed by $^1$H-NMR and $^{13}$C-NMR spectroscopy (Figure 2A, B). The obtained results provide clear evidence for the formation of the imidazolium ring (Figure 2A peaks A, C, D; Figure 2B, peaks A – D). However, the intensity of signal A (at 8.37 ppm in $^1$H-NMR spectrum) is explicitly too low when compared to other peaks. Such behavior is commonly observed in $^1$H-NMR spectra recorded for imidazolium molecules in D$_2$O, which is attributed to the fast exchange of the acidic C2 proton with deuterium atoms of the NMR solvent. In the presence of a sufficiently basic anion such as acetate imidazolium salts exist in equilibrium with the C2-deprotonated carbene and acetic acid,[36,37] which would enhance the proton/deuterium exchange. In our case, this effect was additionally amplified by the harsh conditions (70 °C for 4 h) necessary to dissolve polyC5$^+$Ac$^-$ powder in D$_2$O. The proton-exchange affects also the pattern of its $^{13}$C-NMR spectrum because of the altered chemical environment around carbon A (Figure 2B), which may occur when neighboring protons are replaced by deuterium and broaden its signal. Since in the Debus-Radziszewski reaction the ring closure reaction takes place at the end of several steps by inserting the –CH$_2$– unit into the ring, a quantitative analysis is required here to prove the full closure of the heterocyclic cation rings along the polymer backbone.

Due to limited solubility of freeze-dried polyC5$^+$Ac$^-$ sample in other NMR solvents, anion metathesis reaction was performed (Figure 3) to replace Ac$^-$ in polyC5$^+$Ac$^-$ by bis(trifluromethane sulfonyl)imide anion (TFSI$^-$) (Figure 2C), a large sized, fluorinated anion. This reaction turns the polyimidazolium from being hydrophilic to hydrophobic, thus precipitating out of aqueous solution. The newly formed polymer polyC5$^+$TFSI$^-$ was fairly soluble in DMSO and was investigated by $^1$H-NMR in a non-protic solvent, DMSO-$d_6$ (Figure 2C). This spectrum clearly shows the presence of the C2 proton (Figure 2C, proton A) of imidazolium ring, which has an identical integration value in comparison to the other ring proton (Figure 2C, proton C), *i.e.*, the ring closure reaction indeed takes place and reaches 100%. Therefore the polyC5$^+$Ac$^-$ chemical structure is in consistence with that presented in Figure 1, proving the ring closure reaction during imidazolium structure formation.

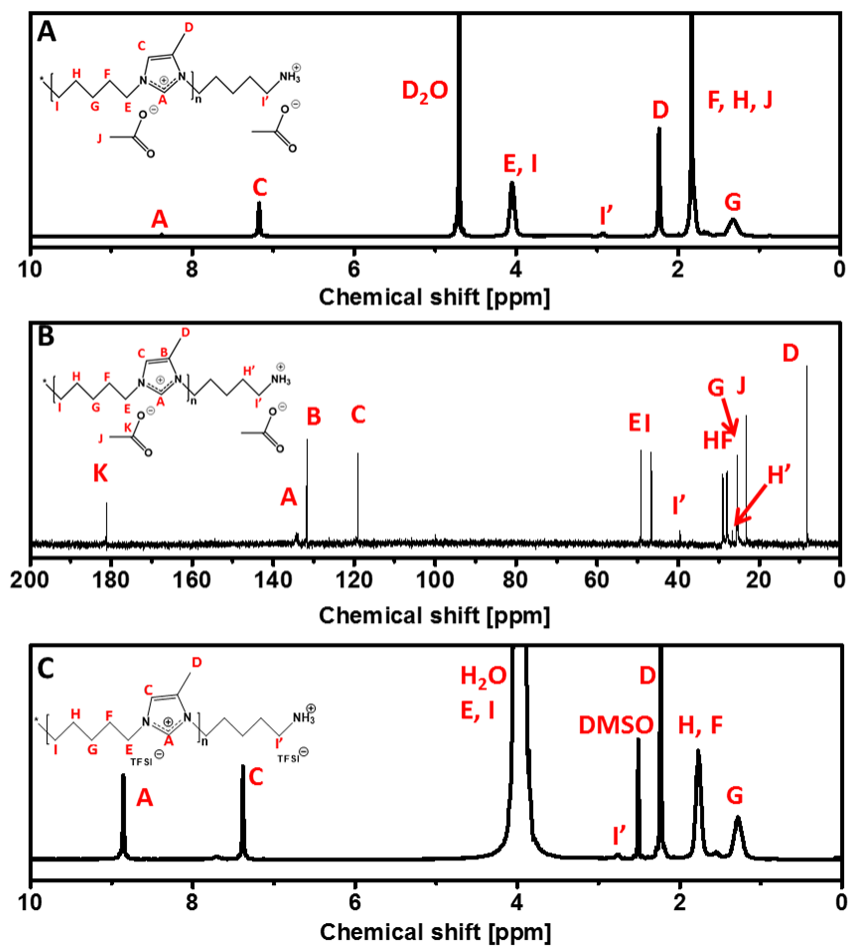

**Figure 2.** A - $^1$H-NMR spectrum of polyC5$^+$Ac$^-$ in D$_2$O; B - $^{13}$C-NMR spectrum of polyC5$^+$Ac$^-$ in D$_2$O; C - $^1$H-NMR of polyC5$^+$TFSI$^-$ in DMSO-$d_6$ (the signal of water in DMSO-$d_6$ is shifted to 3.9 ppm and overlaps with E and I.)

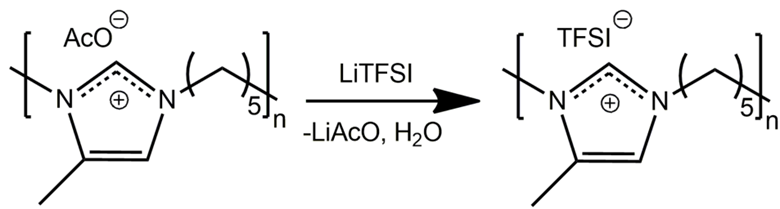

**Figure 3.** Anion metathesis reaction of polyC5$^+$Ac$^-$ with LiTFSI salt in aqueous solution.

It should be noted that during the step-growth polymerization to form polyC5$^+$Ac$^-$, the reaction mixture quickly changes its color from pale yellow to brown after several minutes, which macroscopically indicated the rapid progress of the reaction. In order to quantify the kinetic growth of the polyC5$^+$Ac$^-$ chain over time, kinetic studies were performed *via* gel permeation chromatography (GPC) measurements. We selected a methanol/acetate buffer mixture as an eluent to minimize interactions of the charged polymers with the GPC column. According to the obtained GPC traces upon reaction time, polyC5$^+$Ac$^-$ developed its molar mass quickly within the first hour of the reaction (black and green curves - after 5 min and 1 h of the reaction, respectively, Figure 4) and after that only a slow increase in molar mass of the product can be observed (brown and violet curves – after 8 and 12 h along the reaction time). The reaction reaches its end after 12 h as the GPC trace after that remains essentially unchanged. To determine the absolute molecular weight of polyC5$^+$Ac$^-$, we performed analytical ultracentrifugation (AUC) measurements at different polymer concentrations. AUC experiments confirmed a weight-averaged molecular weight $M_w$ = 24 kg mol$^{-1}$ (± 10 %), corresponding to a degree of polymerization of ~115.

The aforementioned kinetic data clearly indicate that the high molecular weight polymer was formed at room temperature quickly with the majority completed within 1 h. We assume that while the difference in reactivity of butane-1,4-diamine and cadaverine can be neglected, the substitution of glyoxal in Lindner's work with methylglyoxal (pyruvaldehyde) in our case may have a major impact on the kinetics of the reaction due to the electron-donating character of the methyl group, allowing the polymerization to proceed more efficiently. Figure 5 shows the Fourier-transform infrared (ATR-FTIR) spectrum of polyC5$^+$Ac$^-$. Here characteristic peaks of the imidazolium cation ring and the acetate counterion can be clearly identified: the pronounced peaks at 1379 and 1151 cm$^{-1}$ are attributed to C−N and C−C stretching vibrations of the imidazolium ring and the alkyl chains, respectively, while the strong peak at 1568 cm$^{-1}$ is assigned to the C-O stretching of the acetate counteranion. According to the mechanism of the reaction presented by Enthaler *et al.*[27] the reaction goes through the formation of imine, followed by the ring closure reaction. Since in the FTIR spectrum of the product (Figure 5), we do not observe strong bands in the region between 1700 and 1750 cm$^{-1}$ which are characteristic -C=N- stretching bands of the imines [27] we are expecting –NH$_3$OAc as the end groups in the synthesized polymers.

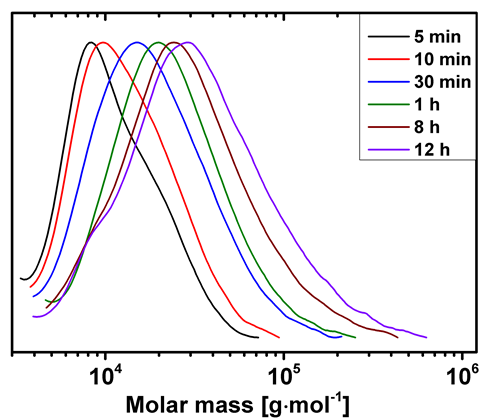

**Figure 4.** GPC traces of polyC5$^+$Ac$^-$ along the polymerization time.

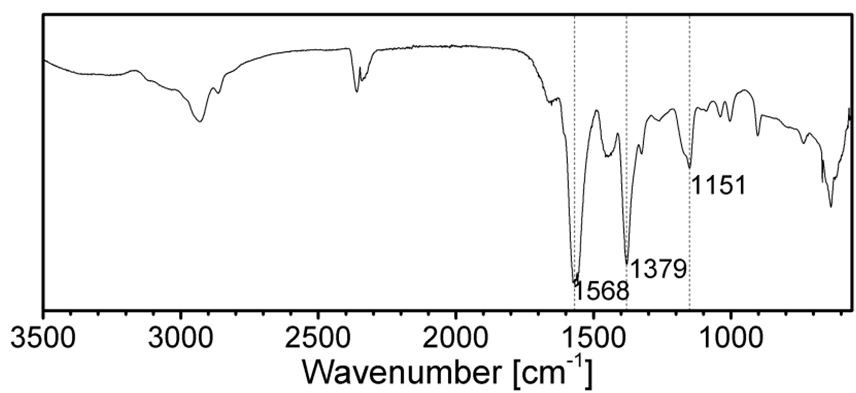

**Figure 5.** ATR/FTIR spectrum of polyC5$^+$Ac$^-$.

To extend the synthetic scope, we performed analogous reactions of pyruvaldehyde, formaldehyde and acetic acid with various diamine compounds. Generally speaking, in comparison with classic approaches to polyimidazoliums, the developed method here simplifies the synthetic procedure and particularly allows for modification of charge density on the polymer backbone by varying the length of the alkyl spacer in the diamine compounds. For this reason, 1,4-tetramethylenediamine, 1,6-hexamethylenediamine, 1,8-octamethylenediamine, 1,10-decamethylenediamine, and 1,12-dodecamethylenediamine were tested for the synthesis of a series of imidazolium polymers (Figure 6). Similar to polyC5$^+$Ac$^-$, their chemical structures were investigated by $^1$H-NMR and FTIR spectroscopy. In the $^1$H-NMR spectra (Figure 7), the only difference among these samples is located in the range from 0 to 2.5 ppm, which is the characteristic chemical shift range for the alkyl chains of different length. In the case of FTIR spectra, the difference is less pronounced (Figure 8).

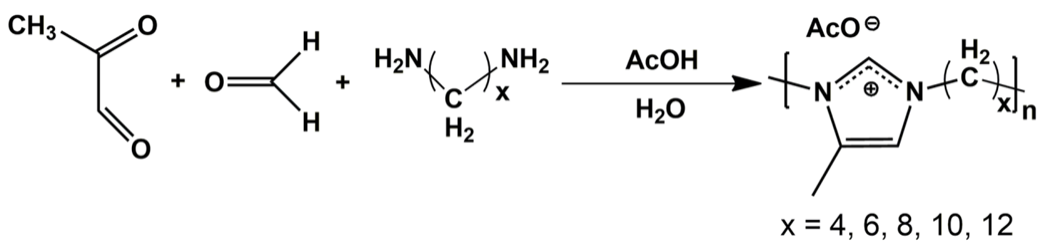

**Figure 6.** The one-pot room temperature synthetic route towards polyimidazolium polymers having different charge densities along their backbones.

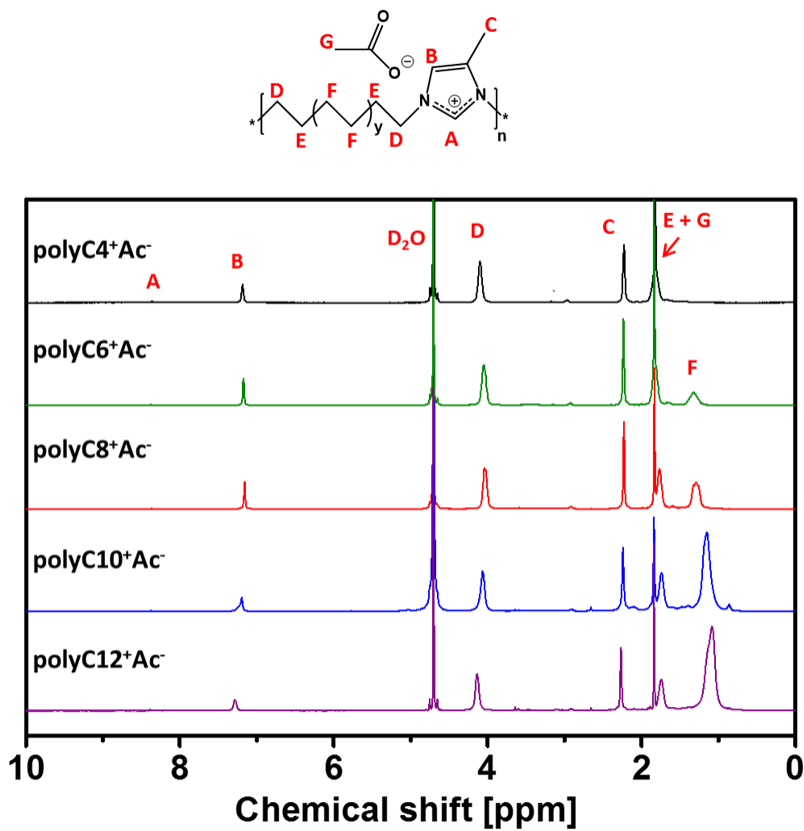

**Figure 7.** $^1$H-NMR spectra of polyCx$^+$Ac$^-$ ($x$ = 4, 6, 8, 10 and 12) recorded in D$_2$O [normalized to the height of the peaks at 4.02 ppm, $y$ – number of –CH$_2$– units containing protons f; in the scheme $y = \frac{X-4}{2}$].

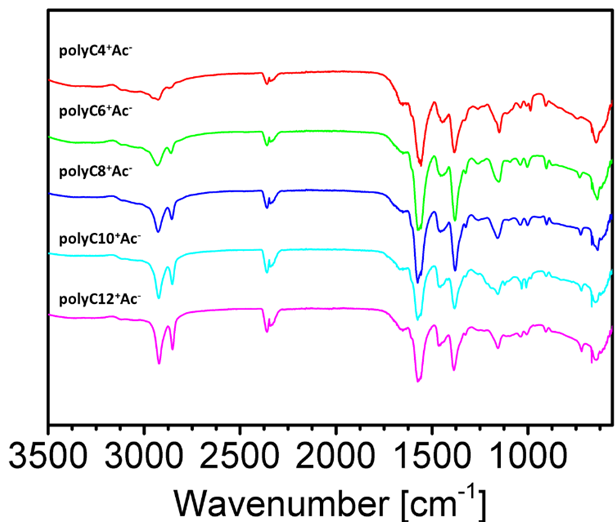

**Figure 8.** ATR/FTIR spectra of polyCx$^+$Ac$^-$ ($x$ = 4, 6, 8, 10 and 12).

### 3.3. Synthesis and Characterization of Imidazolium Polymers From Aromatic Diamines

Motivated by the success in synthesizing main-chain polyimidazoliums bearing tunable charge density through the alkyl spacer in the main chain, we were interested in introducing more rigid aromatic units, such as the phenyl ring into the main chain. As one extreme case, p-phenylenediamine was applied here, aiming at creating a conjugated polyelectrolyte bearing alternative imidazolium and aryl units. The high density of $sp^2$ carbons and –C=N-C– moieties is of particular interest for the generation of nitrogen-doped carbons in a high yield. Figure 9A depicts the synthetic scheme, which involves an anion exchange reaction at the end to replace Ac$^-$ with a nitrogen-rich dicyanamide (DCA) anion. The successful synthesis of polyPh$^+$DCA$^-$ was firstly verified by $^1$H NMR spectroscopy using DMSO-$d_6$ as solvent. As shown in Figure 9B, the proton signals of >C-*CH$_3$*, >C=*CH*-N<, and >N=*CH*-N< on and in the imidazolium ring appear at 2.2, 7.0 and 9.6 ppm, respectively, while the phenyl protons show signals at 7.3 and 7.4 ppm.

FTIR measurement further confirmed the chemical structure of polyPh$^+$DCA$^-$ (Figure 10A). The band at 2145 cm$^{-1}$ is assigned to the characteristic C≡N stretching vibration. Strong absorptions due to C=C/C=N stretching vibrations found at 1508 and 1600 cm$^{-1}$ reveal the presence of imidazolium ring, which is confirmed by the imidazolium ring-breathing mode at 1251 and 1309 cm$^{-1}$. The in-plane C-H deformation of the benzene and imidazolium rings is found in the region of 900−1200 cm$^{-1}$. Additionally, in the range of 800−900 cm$^{-1}$ we can clearly observe the out-of-plane deformations of the substituted benzene rings. Additionally, the polymeric nature of the polyPh$^+$Ac$^-$ sample after dialysis treatment was investigated by GPC. Even though the synthesis was terminated at an early stage (15 min) to access a product that is soluble in water, the apparent number average molecular weight of this polyPh$^+$Ac$^-$ can reach $2 \times 10^4$ g mol$^{-1}$ (Figure 10B). Finally, the high glass transition temperature of the targeted polyPh$^+$DCA$^-$ (~153 °C) provides us additional thermal property information of the obtained polymeric material.

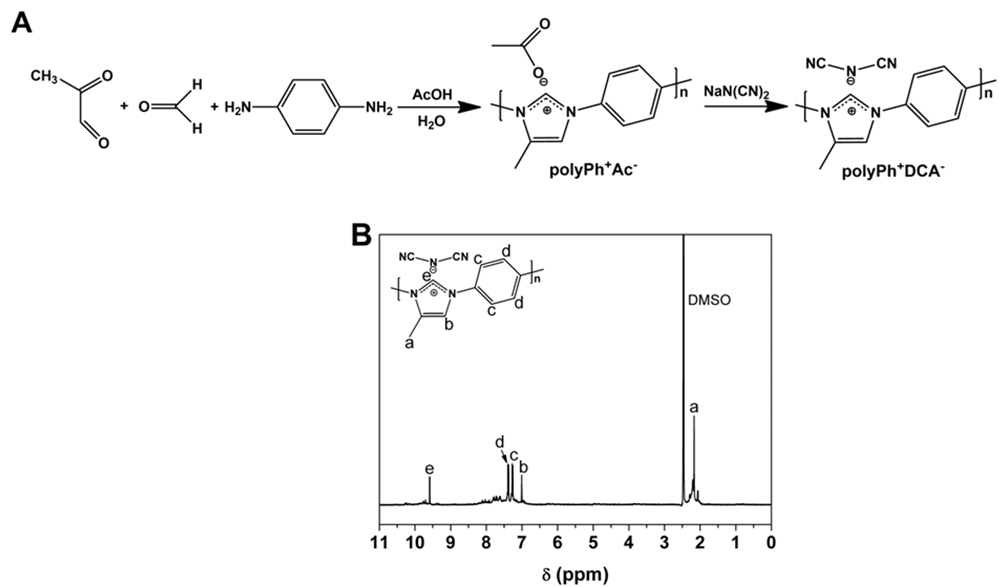

**Figure 9.** A - Schematic illustration of the polyPh$^+$DCA$^-$ synthesis; B - its $^1$H-NMR spectrum.

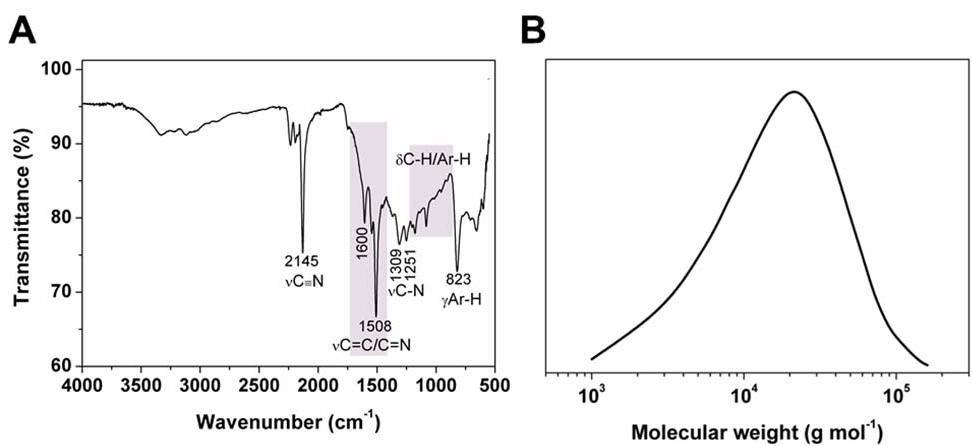

**Figure 10.** A - FTIR spectrum of polyPh$^+$DCA$^-$; B - A GPC trace of polyPh$^+$Ac$^-$.

### 3.4. Synthesis of N-Doped Carbon Materials from the Main-chain Imidazolium Polymers

In order to determine the potential of polyPh$^+$DCA$^-$ as nitrogen-doped carbon precursor, its thermal stability was studied by thermogravimetric analysis (TGA) under nitrogen atmosphere (Figure 11A, B). The slight weight loss of polyPh$^+$DCA$^-$ at 150 °C can be attributed to evaporation of a tiny amount of water. A substantial weight loss was observed in the temperature range between 360 and 510 °C, which is associated with the crosslinking reaction of cyano groups as well as dehydrogenation reaction. Surprisingly, the mass residue at 600 and 900 °C is as high as 72.9 and 66.2 wt %, respectively. A real carbonization experiment performed in an oven gave a carbonization yield of 66.8 wt %, in good agreement with the above mentioned TGA analysis. The presence of inorganic impurities such as sodium salt that was used in the anion exchange process was excluded (below 0.05 wt %) by inductively coupled plasma-optical emission spectroscopy. The carbon and nitrogen contents of the carbon product at 900 °C are determined to be 80.7 and 10.7 wt %. We assume that the mass residue of polyPh$^+$DCA$^-$ at 900 °C, which is among the highest ones in synthetic polymers,[38-41] derives from the judicious combination of chemical units, such as the cyano group, the aromatic phenyl group and the imidazolium ring, which can favorably end up in the final carbon matrix with little-to-no weight loss.

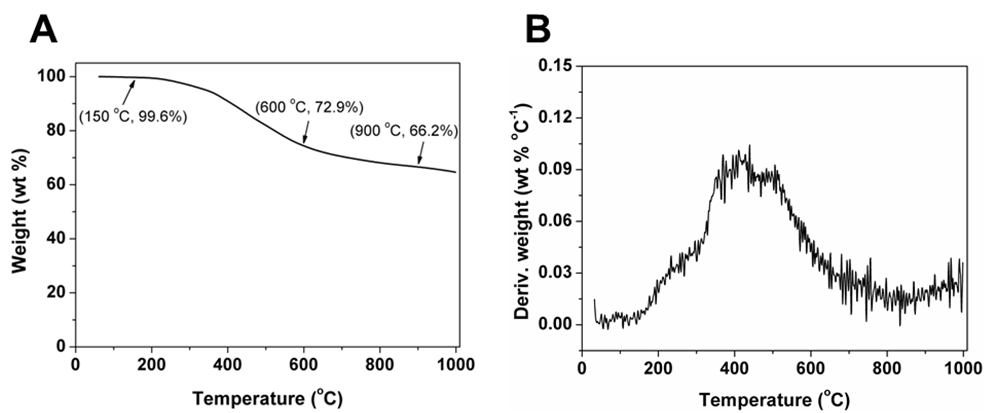

**Figure 11.** A - TGA curve of polyPh$^+$DCA$^-$ (the measurement performed under nitrogen atmosphere); B - DTG (the first derivative of TGA curve) curve for polyPh$^+$DCA$^-$ under nitrogen atmosphere at 10 °C min$^{-1}$.

XPS measurement was conducted to further analyze the surface element composition and binding motif of the obtained nitrogen-doped carbon. Figure 12A shows the survey scan XPS spectrum with apparent C 1s (284.6 eV), N 1s (398.6 eV) and O 1s (532.3 eV) peaks. The high-resolution N 1s XPS spectra are curve-fitted into four individual peaks: pyridinic N (398.2 eV), pyrrolic N (399.7 eV), graphitic N (400.7 eV), and oxidized N (402.8 eV), which follows our previous work.[39] As displayed in Figure 12B, the quantitative analyses indicate that pyridinic N (49.9–51.7%) and graphitic N (29.8–31.3%) are the two most abundant N bonding schemes in the resultant nitrogen-doped carbon, as typical for nitrogen-doped carbons prepared from nitrogen-rich molecules at high temperatures.[42]

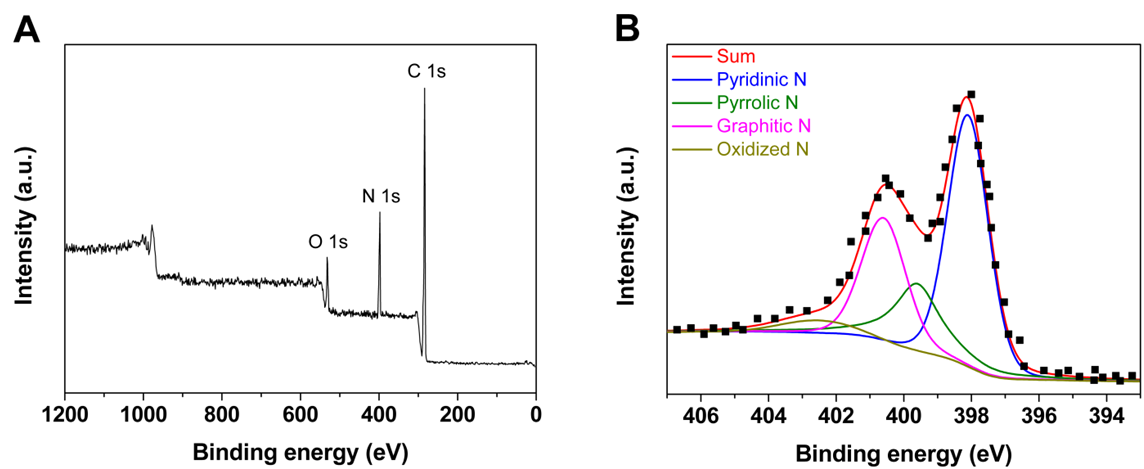

**Figure 12.** A – XPS curve of the nitrogen-doped carbon from the carbonization of polyPh$^+$DCA$^-$ at 900 $^o$C; B – High-resolution N 1s XPS curve of the nitrogen-doped carbon.

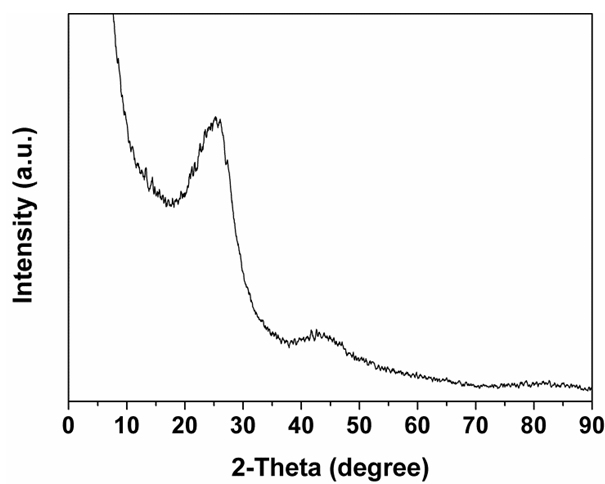

**Figure 13.** XRD pattern of the nitrogen-doped carbon.

XRD measurement was employed to characterize the local order of the nitrogen-doped carbon. As depicted in Figure 13, two broad diffraction peak at $2\theta = 26.1^o$ and $44^o$ are assigned to the typical graphitic (002) and (101) planes,[43-45] while a weak but still recognizable peak at about 80$^o$, assigned to the (110) plane, which confirms a restricted stacking of the graphitic sheets.

## 4. Conclusions

In conclusion, we demonstrated a one-pot synthesis of main-chain imidazolium polymers which employs the Debus-Radziszewski reaction to build up the polymer main chain. The presented synthesis is simple, high-yielding and also sustainable, as the presented reaction was performed in aqueous media at ambient conditions in an energy- and time-efficient fashion. This elegant and synthetically convenient protocol allows us to produce polymeric materials partially or exclusively from small, bio-mass derived compounds. We also confirmed that this synthetic approach can be used to easily tune the structure and properties of the synthesized materials towards a variety of ionic polymers. A very high carbonization yield of 66.8 wt % was achieved in the carbonization of one of these materials derived from p-phenylenediamine and dicyanamide. Thus, in our opinion, the Debus-Radziszewski polymerization and polymer post modifications will find more regular applications, as it is perfect tool for the straightforward synthesis of functional polyelectrolytes.

Acknowledgements:
The authors acknowledge financial support from the Max Planck Society, the Marie Curie Actions of EU's 7[th] Framework Program under REA grant agreement no. 289347, and European Research Council (ERC) starting grant NAPOLI-639720.

Received: Month XX, XXXX; Revised: Month XX, XXXX; Published online:

Keywords: polyimidazolium, sustainable synthesis, carbonization, Debus-Radziszewski reactions